\documentclass{PoS}
\usepackage{mathptmx}
\usepackage{calrsfs}
\usepackage{pdfpages}
\DeclareMathAlphabet{\mathcal}{OMS}{zplm}{m}{n}

\let\OLDthebibliography\thebibliography
\renewcommand\thebibliography[1]{
  \OLDthebibliography{#1}
  \setlength{\parskip}{0pt}
  \setlength{\itemsep}{-3pt}
\footnotesize
}

\title{Theoretical analysis of double parton scatterings in quarkonium production in proton-proton collisions at the LHC}

\ShortTitle{DPS in quarkonium production in $pp$ collisions at the LHC}

\author{\speaker{Nodoka~Yamanaka}\thanks{Supported by JSPS Postdoctoral Fellowships for Research Abroad.}\\
        IPNO, CNRS-IN2P3, Univ. Paris-Sud, Universit\'e Paris-Saclay, 
91406 Orsay Cedex, France\\
        E-mail: \email{yamanaka@ipno.in2p3.fr}
}

\author{Jean-Philippe Lansberg\\
        IPNO, CNRS-IN2P3, Univ. Paris-Sud, Universit\'e Paris-Saclay, 
91406 Orsay Cedex, France\\
}

\author{Hua-Sheng Shao\\
Laboratoire de Physique Th\'eorique et Hautes Energies (LPTHE), UMR 7589, Sorbonne Universit\'e et CNRS, 4 place Jussieu, 75252 Paris Cedex 05, France\\
}

\author{Yu-Jie Zhang\\
        Beijing Key Laboratory of Advanced Nuclear Energy Materials and Physics, and School of Physics, Beihang University, Beijing 100191, China\\
}

\abstract{
The production process of quarkonia in proton-proton ($pp$) collision is a very good probe of the parton structure of the proton.
Recent experimental data of the production of $J/\psi$+vector boson or quarkonium pairs at the LHC and Tevatron suggest the relevance of double parton scatterings (DPS).
We discuss here the single parton scattering (SPS) contribution to the $J/\psi +Z$, $J/\psi +W$, and $J/\psi +J/\psi$ productions in hadron collisions.
By revisiting the computations of the SPS contributions to the $J/\psi+Z$ and $J/\psi +W$ productions, we demonstrate that the ATLAS data in fact show evidence for DPS.
}

\FullConference{XXVI International Workshop on Deep-Inelastic Scattering and Related Subjects (DIS2018)\\
		16-20 April 2018\\
		Kobe, Japan}

\begin{document}

\section{Introduction}

The study of quarkonium production at colliders can be used to probe perturbative and nonperturbative properties of QCD.
Indeed, the production of $J/\psi +W$ was proposed as a golden channel to probe the color-octet contribution and thus to test the nonrelativistic QCD (NRQCD) \cite{Barger:1995vx}.
It is also interesting in the point-of-view of the search for new physics beyond the standard model.
The $\Upsilon +W$ production process could be a decay channel of a charged Higgs boson \cite{Grifols:1987iq}.
The associated production of a quarkonium and a photon was proposed to constrain the quarkonium-production 
mechanisms~\cite{Roy:1994vb,Mathews:1999ye,Li:2008ym,Lansberg:2009db,Li:2014ava} and to study the gluon distribution in the proton~\cite{Doncheski:1993dj,Dunnen:2014eta}.
On the other hand, the $J/\psi +J/\psi$ production could be a key process to study double parton scatterings (DPS) \cite{Kom:2011bd,Baranov:2015cle,Borschensky:2016nkv,Lansberg:2014swa} and to look for linearly polarized gluons in the proton~\cite{Lansberg:2017dzg}.
Accurate studies of the DPS are also important in the context of beyond the standard model physics, since it may be a significant background in multi-particle final states of high-energy $pp$ collisions.

The experimental study of the quarkonium associated production recently experienced much progress.
The ATLAS Collaboration observed the $J/\psi +W$ \cite{Aad:2014rua} and $J/\psi +Z$ \cite{Aad:2014kba} final states.
The experimental data of $J/\psi +J/\psi$ production was also studied by many experiments, such as D0 \cite{Abazov:2014qba}, 
CMS \cite{Khachatryan:2014iia}, ATLAS \cite{Aaboud:2016fzt}, and LHCb \cite{Aaij:2011yc,Aaij:2016bqq} Collaborations. 
On the other hand, theoretical computations of $J/\psi+W$ and $J/\psi+Z$ via single-parton scatterings (SPS) were carried out up to NLO in $\alpha_s$ \cite{Li:2010hc,Mao:2011kf,Gong:2012ah,Lansberg:2013wva} 
whereas only partial NLO NRQCD contribution exists for $J/\psi+J/\psi$.

In this proceedings contribution, we report on the SPS contribution to the $J/\psi +W$, $J/\psi +Z$, and $J/\psi +J/\psi$ \cite{Lansberg:2014swa,Lansberg:2013qka,He:2015qya,Sun:2014gca} productions in the color evaporation model (CEM).

\section{Analysis of ATLAS data for $J/\psi +Z$ and $J/\psi +W$ productions in the CEM}

\begin{figure}[hbt!]
\begin{center}
\includegraphics[width=1.0\columnwidth]{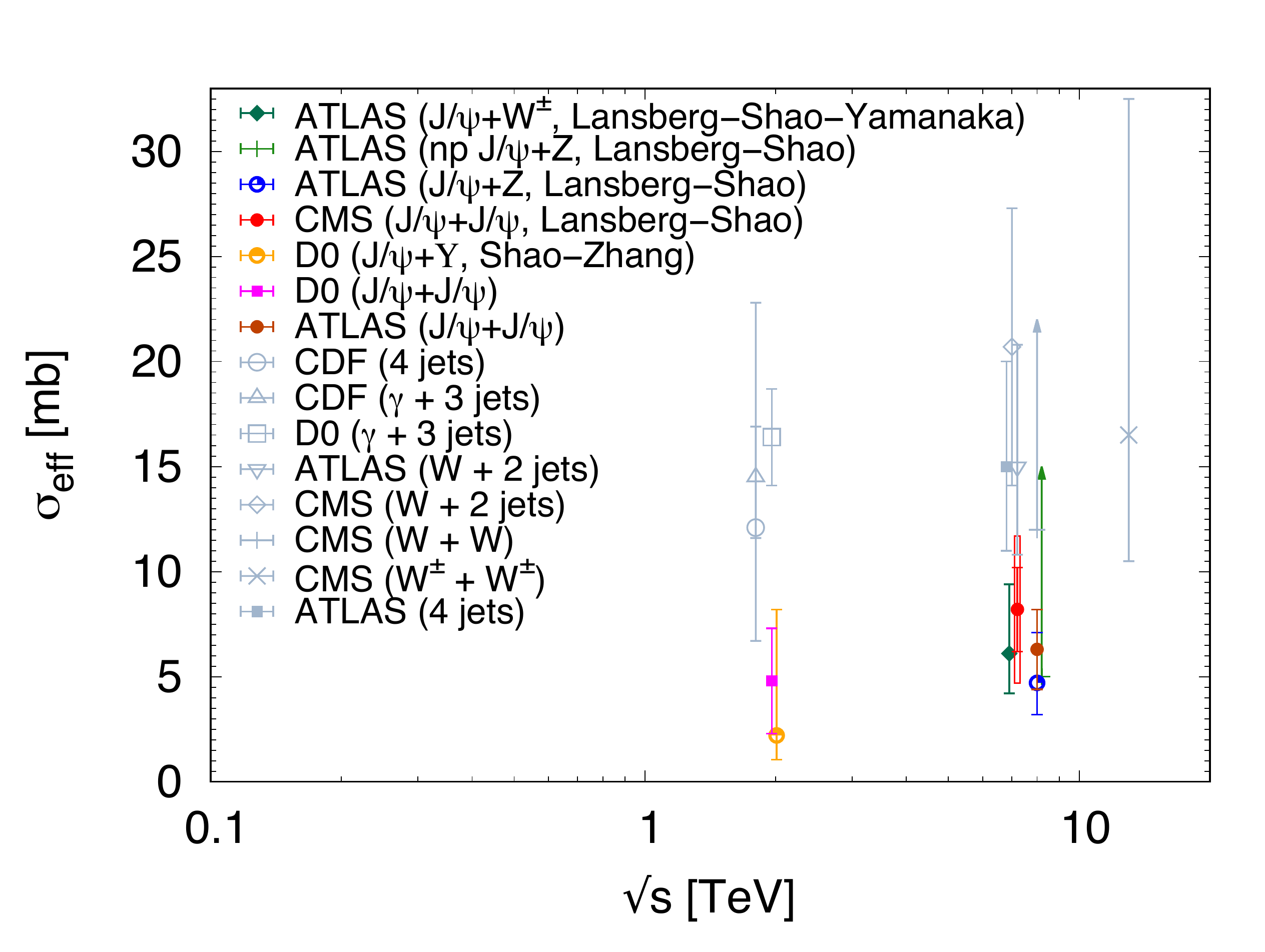}
\caption{
Comparison of $\sigma_{\rm eff}$ extracted by several experiments and theoretical calculations from the $J/\psi +Z/W$ and double quarkonium productions \cite{Lansberg:2017chq}.
Quarkonium related extractions are in color.
}
\label{fig:sigmaeff_summary}
\end{center}
\end{figure}

At high energies, multiple parton interactions can become relevant, despite of being formally higher twist.
They are in fact necessary to restore the unitarity of the cross section and are enhanced by the strong increase of the parton density at high energies.
Double hard parton scatterings fall in this category.
Assuming that both parton collisions occur independently, one usually parametrizes the DPS cross sections by the so-called pocket-formula:
\begin{equation}
\sigma_{\rm DPS} (A+B)
=
\frac{1}{1+\delta_{AB}}
\frac{\sigma (A) \sigma (B)}{\sigma_{\rm eff}}
,
\end{equation}
where $\delta_{AB} =1$ for the $J/\psi+J/\psi$ final state and $\delta_{AB} =0$ for the $J/\psi+W/Z$ ones.
We plot in Fig. \ref{fig:sigmaeff_summary} the current situation of the extractions of $\sigma_{\rm eff}$.

\begin{table}[htb]
\caption{
Comparison of the experimental data of ATLAS with several theoretical results.
}
\begin{tabular}{l|llll}
\hline
& ATLAS & DPS ($\sigma_{\rm eff} = 15$ mb) & CSM & COM \\
\hline
$J/\psi +Z$ & $1.6 \pm 0.4$ pb \cite{Aad:2014kba} & $0.46$ pb & $0.025 - 0.125$ pb \cite{Gong:2012ah} & $< 0.1$ pb \cite{Mao:2011kf} \\
$J/\psi +W$ & $4.5 ^{+1.9}_{-1.5}$ pb \cite{Aad:2014rua} & $1.7$ pb & $(0.11 \pm 0.04)$ pb \cite{Lansberg:2013wva} & $(0.16 -0.22)$ pb \cite{Li:2010hc} \\
\hline
\end{tabular}
\label{table:comparison}
\end{table}

Let us compare the experimental data of ATLAS for the $J/\psi +W$ and $J/\psi +Z$ productions with several results of theoretical calculations.
The comparison is shown in Table \ref{table:comparison}.
We see that the results of ATLAS are significantly above the SPS contribution (color singlet model and color-octet mechanism, abbreviated as CSM and COM, respectively), and the DPS with $\sigma_{\rm eff}$, determined by the ATLAS $W+$ 2jets data ($\sigma_{\rm eff} = 15$ mb).
They can only account for a fraction of the data (deviations of $>3 \sigma$ for $J/\psi +Z$, $>2 \sigma$ for $J/\psi +W$).
A natural question then arises: is the SPS underestimated?

To estimate the upper limit of the SPS, we use the CEM.
In the CEM, the quarkonium final state is formed when the invariant mass of the heavy quark pair remains below the open-heavy flavor threshold, and the cross section is then derived from 
\begin{equation}
\sigma^{\rm (N)LO,\ \frac{direct}{prompt}}_{ J/\psi}
=
{\cal P}^{\rm (N)LO,\frac{direct}{prompt}}_{J/\psi}\int_{2m_c}^{2m_D} 
\frac{d\sigma_{c\bar c}^{\rm (N)LO}}{d m_{c\bar c}}d m_{c\bar c}
,
\label{eq:sigma_CEM}
\end{equation}
where ${\cal P}^{\rm (N)LO,{prompt}}_{J/\psi} = 0.014$ (LO), 0.009 (NLO) \cite{Lansberg:2016rcx} is expected to be nonperturbative but universal. 
The single-quarkonium production in the CEM overshoots the experimental data at high transverse momentum $p_T$ \cite{Lansberg:2006dh,Andronic:2015wma,Lansberg:2016rcx}.
This is due to the dominance of the gluon fragmentation.
The same phenomenon is expected to occur for $J/\psi +W$ and $J/\psi +Z$ productions.
The CEM gives us a conservative upper limit on the SPS yield.
We compute it in both cases at NLO with {\small \sc MadGraph5\_aMC@NLO} \cite{Alwall:2014hca}.

\begin{figure}[htb]
\begin{center}
\includegraphics[width=.49\columnwidth]{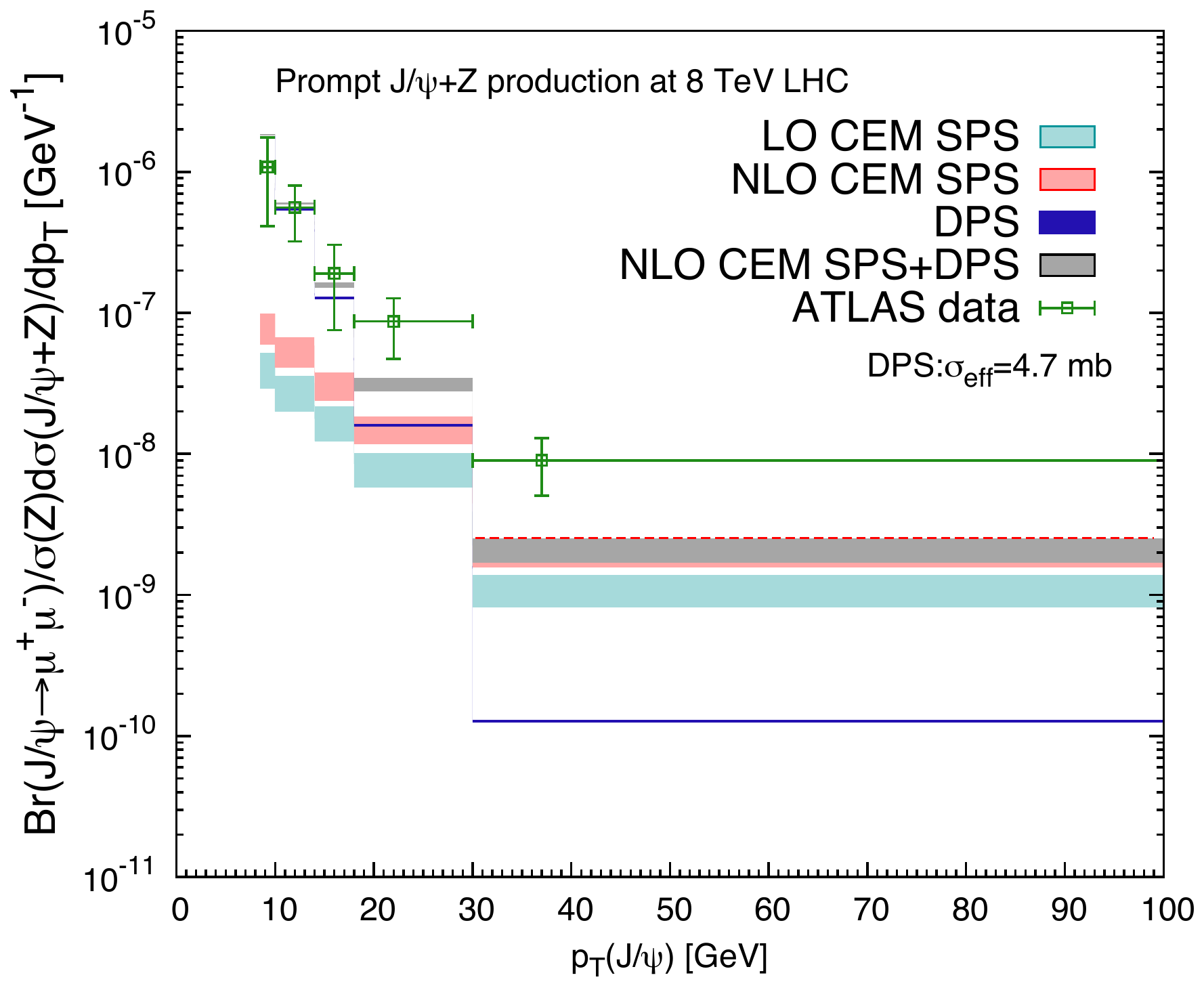}
\includegraphics[width=.49\columnwidth]{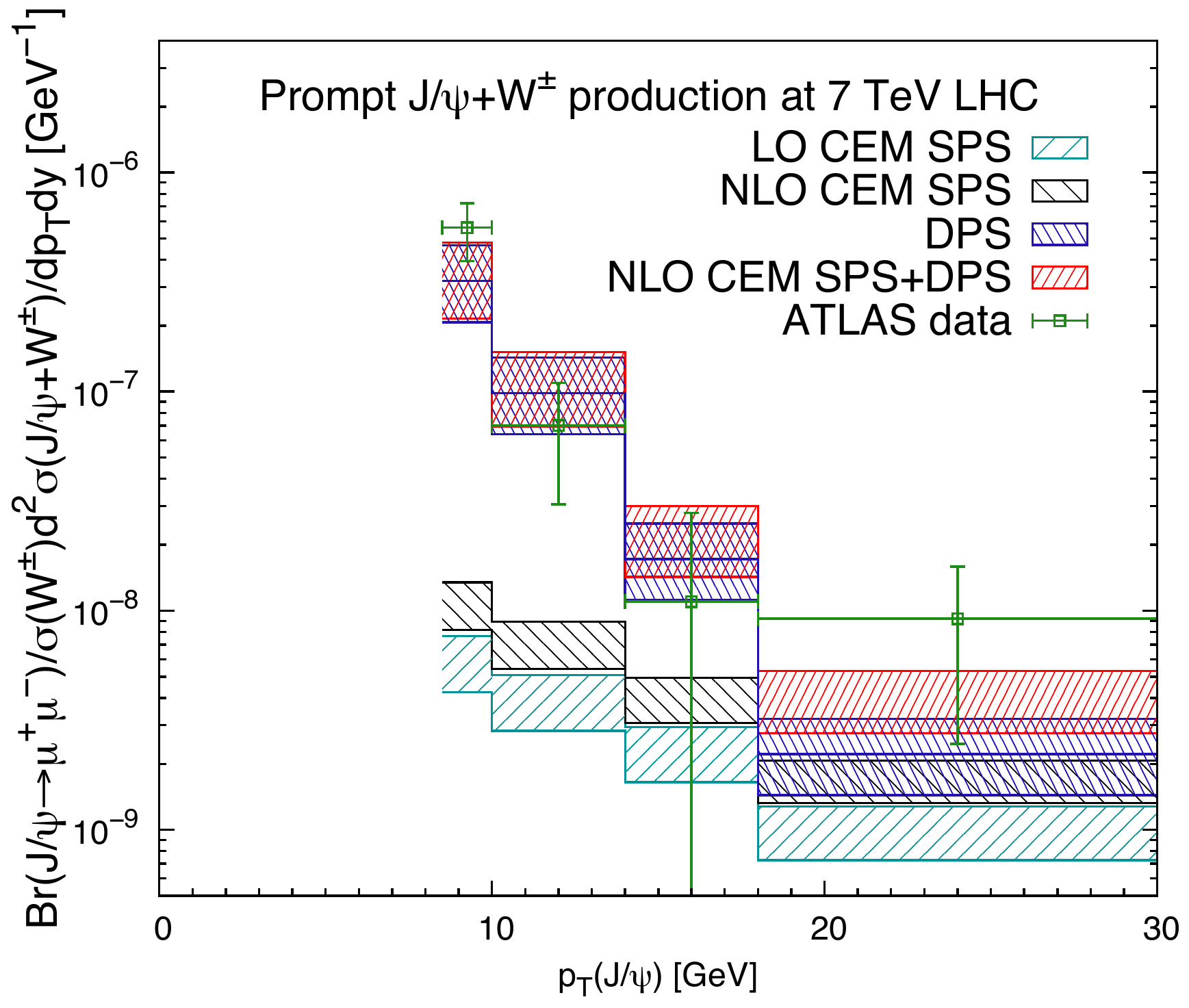}
\caption{
The $p_T$ dependence of the $J/\psi$ in the $J/\psi +Z$ \cite{Lansberg:2016rcx} and $J/\psi +W$ \cite{Lansberg:2017chq} production cross section calculated in the CEM.
The experimental data of the ATLAS Collaboration \cite{Aad:2014kba,Aad:2014rua} are also shown.
}
\label{fig:pT_distribution}
\end{center}
\end{figure}

We now show the results for the $J/\psi +Z$ and $J/\psi +W$ productions.
From the NLO CEM calculation, we have \cite{Lansberg:2016rcx,Lansberg:2017chq}
\begin{eqnarray}
\sigma_{J/\psi +Z}
&=&
0.19^{+0.05}_{-0.04}\, {\rm pb}
,
\\
\sigma_{J/\psi +W}
&=&
0.28 \pm 0.07\, {\rm pb}
,
\end{eqnarray}
where the error bars are the combined statistical and systematic uncertainties.
We see that the upper limits by the CEM alone do not solve the discrepancy between the SPS and the measurements in particular at low $p_T$ (also compare with the numbers in Table \ref{table:comparison}).

\begin{figure}[hbt]
\begin{center}
\includegraphics[width=.46\columnwidth]{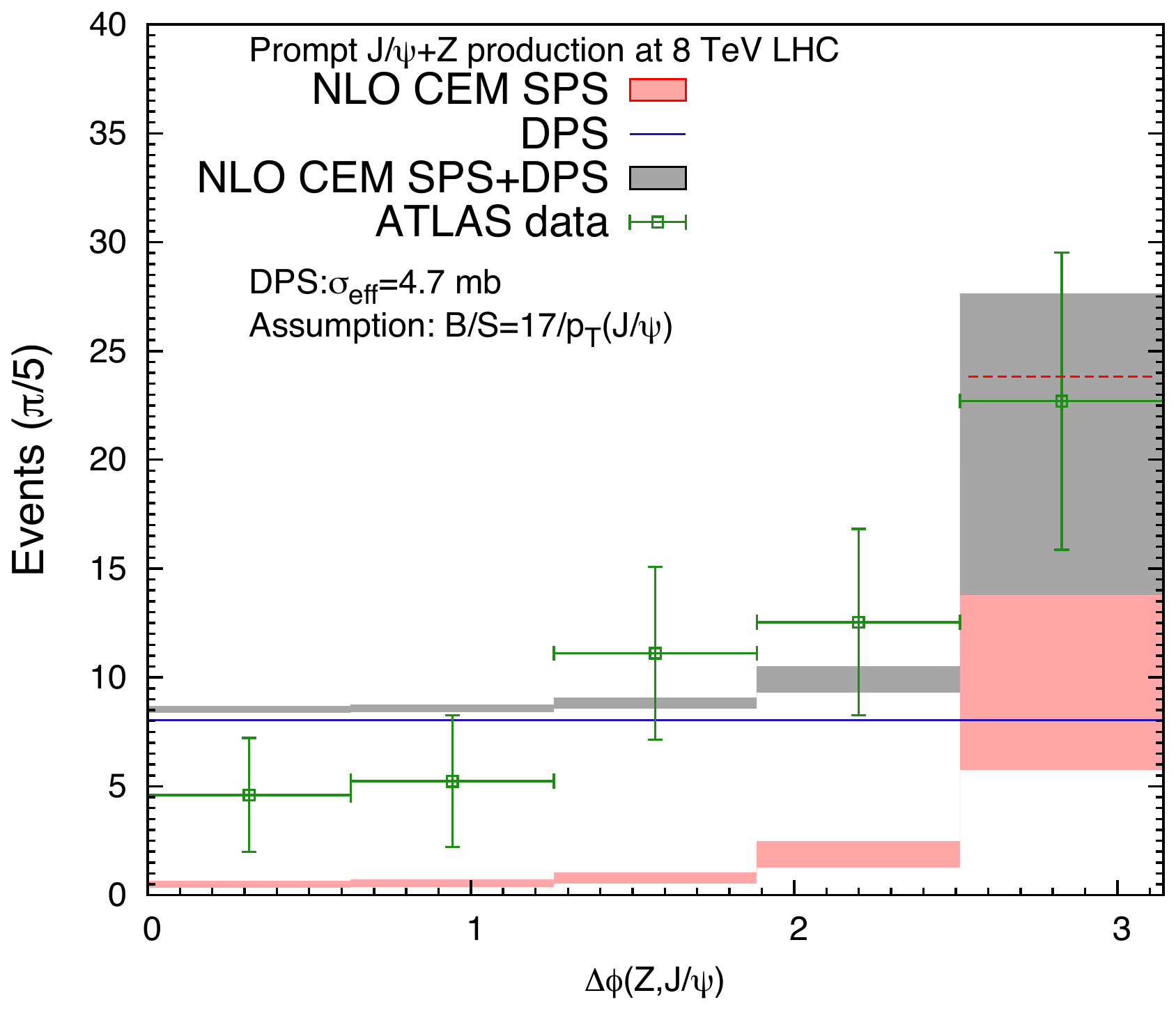}
\includegraphics[width=.46\columnwidth]{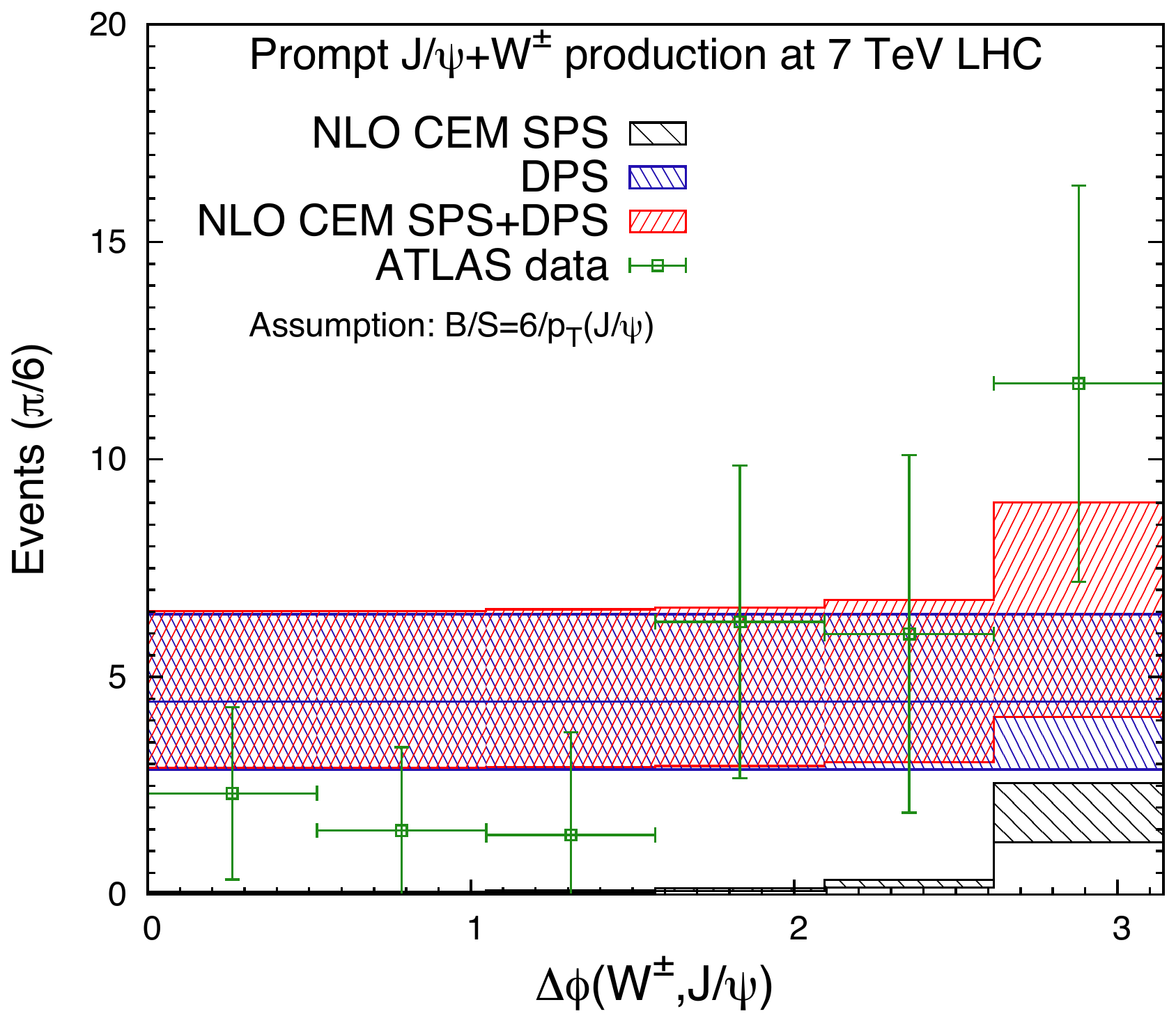}
\caption{
The event distribution in $\Delta \phi$ of the $J/\psi +Z$ \cite{Lansberg:2016rcx} and $J/\psi +W$ \cite{Lansberg:2017chq} production cross section calculated in the CEM.
The experimental data of ATLAS Collaboration \cite{Aad:2014kba,Aad:2014rua} are also shown.
}
\label{fig:angular_distribution}
\end{center}
\end{figure}

Let us now see whether this gap disappears by increasing the DPS.
We fit $\sigma_{\rm eff}$ to the ATLAS data with the SPS contribution subtracted.
The results for the $p_T$ differential cross section are shown in Fig. \ref{fig:pT_distribution}.
By fitting the difference between the experimental data of inclusive total cross section measurements and the CEM predictions, we obtain $\sigma_{\rm eff} = (4.7 ^{+2.4}_{-1.5})$ mb for the ${J/\psi +Z}$ production \cite{Lansberg:2016rcx}, and $\sigma_{\rm eff} = (6.1 ^{+3.3}_{-1.9})$ mb for that of ${J/\psi +W}$ \cite{Lansberg:2017chq}, which are in agreement with each other.
Increasing the DPS seems to solve the puzzle.
We note that the SPS yield favored by the ATLAS acceptance remains visible at $\Delta \phi = \pi$ (see Fig. \ref{fig:angular_distribution}) in the uncorrected $\Delta \phi$ distributions. 

\section{$J/\psi + J/\psi $ in the CEM}

The $J/\psi +J/\psi$ production was measured by CMS \cite{Khachatryan:2014iia}, D0 \cite{Abazov:2014qba}, ATLAS \cite{Aaboud:2016fzt}, and LHCb \cite{Aaij:2016bqq} Collaborations.
The extraction of the DPS contribution was only performed by D0 ($\sigma_{\rm eff} = (4.8 \pm 0.5_{\rm stat} \pm 2.5_{\rm sys})$ mb) and ATLAS ($\sigma_{\rm eff} = (6.3 \pm 1.6_{\rm stat} \pm 1.0_{\rm sys})$ mb).
In Ref. \cite{Lansberg:2014swa}, $\sigma_{\rm eff} = (8.2 \pm 2.0_{\rm stat} \pm 2.9_{\rm sys})$ mb was extracted from the experimental data of CMS based on the NLO$^\star$ SPS calculations~\cite{Lansberg:2013qka} in CSM with the help of {\sc\small HELAC-Onia}~\cite{Shao:2012iz,Shao:2015vga}.
There is however an on-going discussion about the actual size of the SPS in NRQCD \cite{He:2015qya}.
The LO COM yield depends on the square of nonpeturbative color-octet long-distance matrix elements (LDMEs) in NRQCD and is thus affected by large uncertainties. 

\begin{figure}[htb]
\begin{center}
\includegraphics[width=.49\columnwidth]{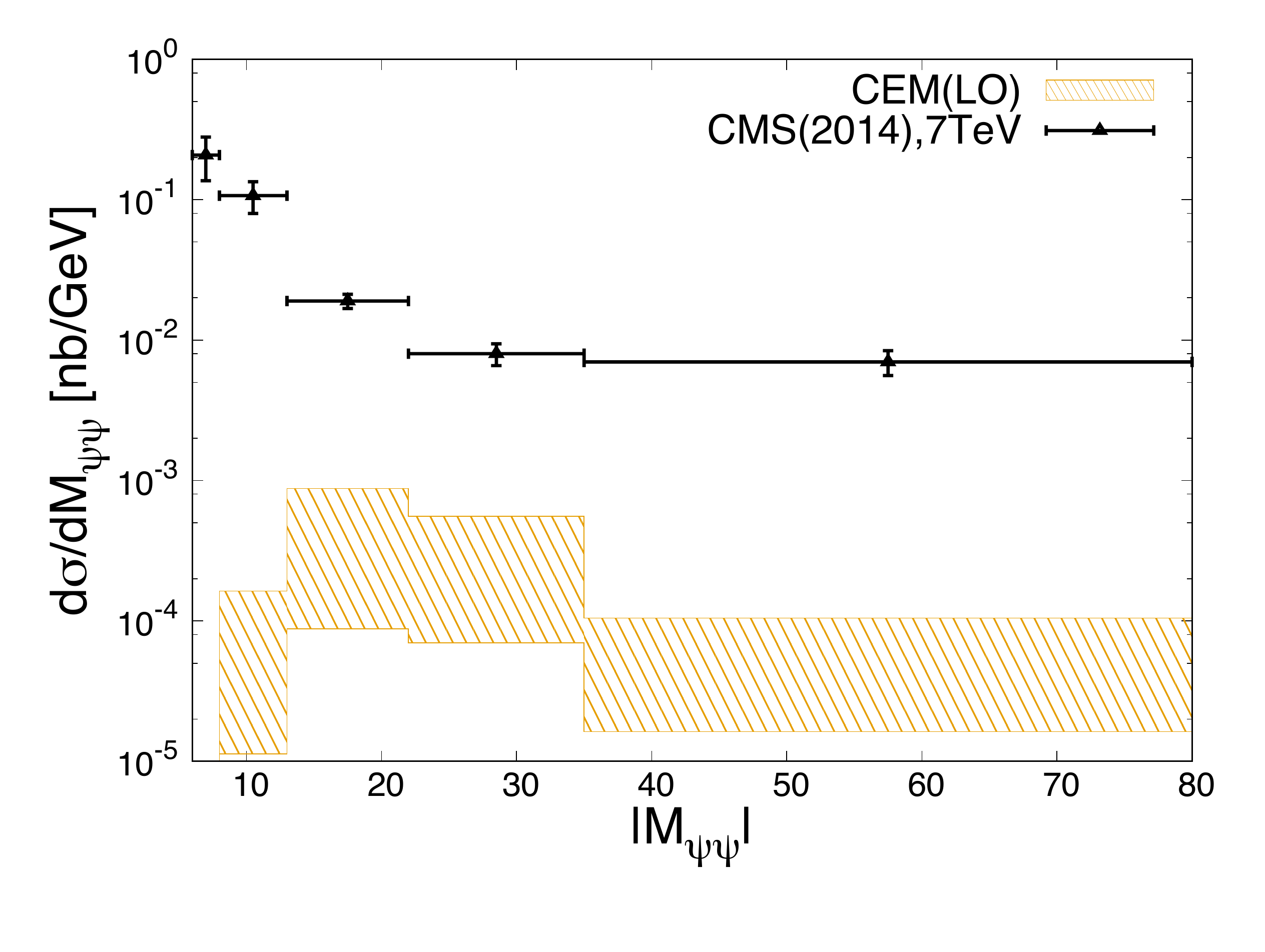}
\includegraphics[width=.49\columnwidth]{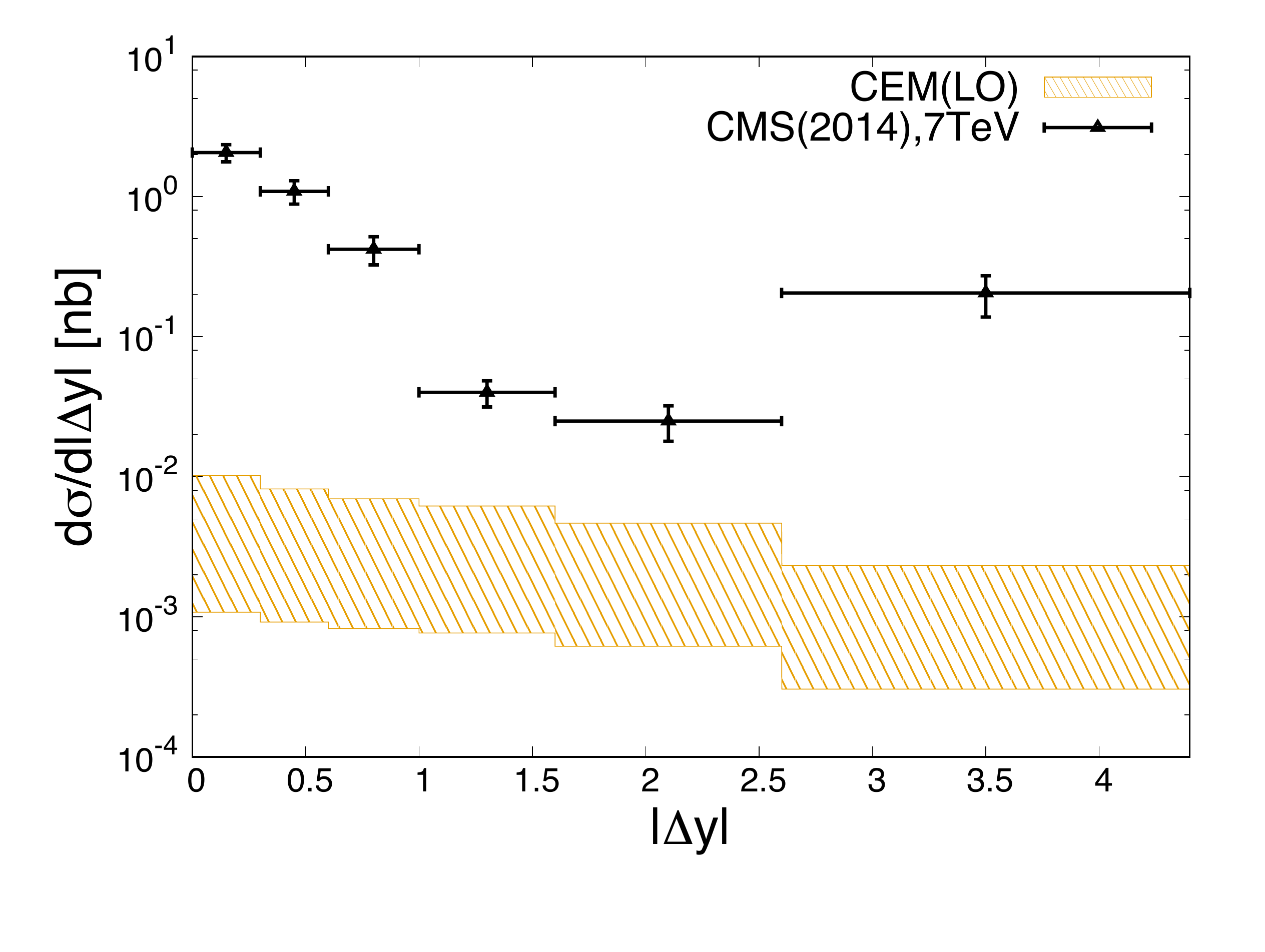}
\caption{
The invariant mass (left panel) and $\Delta y$ (right panel) distributions of $J/\psi$-pair production with the center mass energy 7 TeV (CMS setup).
}
\label{fig:double_Jpsi_CMS7TeV}
\end{center}
\end{figure}

To get the order of magnitude of the contribution from the color-octet transitions, we evaluate the $J/\psi +J/\psi$ production in the CEM at LO.
The CEM yield should give another indication of what the SPS contributions could be at large invariant mass and $\Delta y$.
The result is displayed in Fig. \ref{fig:double_Jpsi_CMS7TeV}.
We note that the CEM is lower than the LO NRQCD result of Ref. \cite{He:2015qya}.
As such their result may indeed be optimistic, with an debatable choice of the LO LDMEs.
This reinforces our confidence in our extraction made in Ref. \cite{Lansberg:2014swa} of $\sigma_{\rm eff} =8$ mb.

\section{Conclusion}

In summary, we studied the production processes of $J/\psi + W/Z$ at the NLO and $J/\psi + J/\psi$ at the LO in $\alpha_s$, relying on a quark-hadron duality.
The associated production of $J/\psi + W/Z$ was measured by ATLAS, and a gap between the experimental data and a SPS+DPS estimation ($\sigma_{\rm eff} = 15$ mb) was seen.
In order to check whether the SPS was underestimated, we evaluated the NLO CEM yields of $J/\psi + W/Z$.
We found that the conservative upper limits set by the CEM do not solve the discrepancy between the ATLAS data and the SPS with a DPS evaluated with  $\sigma_{\rm eff} = 15$ mb.
By fitting $\sigma_{\rm eff}$, we obtained $\sigma_{\rm eff} = (4.7 ^{+2.4}_{-1.5})$ mb (${J/\psi +Z}$), and $\sigma_{\rm eff} = (6.1 ^{+3.3}_{-1.9})$ mb (${J/\psi +W}$).
In fact, $J/\psi + W/Z$ shows evidence for DPS.
Fig. \ref{fig:sigmaeff_summary} summarizes the current status of $\sigma_{\rm eff}$.
All the central rapidity quarkonium data are compatible with a small $\sigma_{\rm eff}$.
The $J/\psi +J/\psi$ production also requires the DPS contribution with $\sigma_{\rm eff}<10$ mb at large invariant mass and $\Delta y$.
Overall, $\sigma_{\rm eff}$ seems to be smaller for centrally-produced quarkonia than for jets, which is maybe a hint for the flavor dependence of DPS.

\bibliographystyle{Science}

\bibliography{yamanaka}

\begin{thebibliography}{10}

\bibitem{Barger:1995vx}
V.~D. Barger, S.~Fleming, R.~J.~N. Phillips, {\it Phys. Lett.\/} {\bf B371},
  111 (1996).

\bibitem{Grifols:1987iq}
J.~A. Grifols, J.~F. Gunion, A.~Mendez, {\it Phys. Lett.\/} {\bf B197}, 266
  (1987).

\bibitem{Roy:1994vb}
D.~P. Roy, K.~Sridhar, {\it Phys. Lett.\/} {\bf B341}, 413 (1995).

\bibitem{Mathews:1999ye}
P.~Mathews, K.~Sridhar, R.~Basu, {\it Phys. Rev.\/} {\bf D60}, 014009 (1999).

\bibitem{Li:2008ym}
R.~Li, J.-X. Wang, {\it Phys. Lett.\/} {\bf B672}, 51 (2009).

\bibitem{Lansberg:2009db}
J.~P. Lansberg, {\it Phys. Lett.\/} {\bf B679}, 340 (2009).

\bibitem{Li:2014ava}
R.~Li, J.-X. Wang, {\it Phys. Rev.\/} {\bf D89}, 114018 (2014).

\bibitem{Doncheski:1993dj}
M.~A. Doncheski, C.~S. Kim, {\it Phys. Rev.\/} {\bf D49}, 4463 (1994).

\bibitem{Dunnen:2014eta}
W.~J. den Dunnen, J.~P. Lansberg, C.~Pisano, M.~Schlegel, {\it Phys. Rev.
  Lett.\/} {\bf 112}, 212001 (2014).

\bibitem{Kom:2011bd}
C.~H. Kom, A.~Kulesza, W.~J. Stirling, {\it Phys. Rev. Lett.\/} {\bf 107},
  082002 (2011).

\bibitem{Baranov:2015cle}
S.~P. Baranov, A.~H. Rezaeian, {\it Phys. Rev.\/} {\bf D93}, 114011 (2016).

\bibitem{Borschensky:2016nkv}
C.~Borschensky, A.~Kulesza, {\it Phys. Rev.\/} {\bf D95}, 034029 (2017).

\bibitem{Lansberg:2014swa}
J.-P. Lansberg, H.-S. Shao, {\it Phys. Lett.\/} {\bf B751}, 479 (2015).

\bibitem{Lansberg:2017dzg}
J.-P. Lansberg, C.~Pisano, F.~Scarpa, M.~Schlegel, {\it Phys. Lett.\/} {\bf
  B784}, 217 (2018).

\bibitem{Aad:2014rua}
G.~Aad, {\it et~al.\/}, {\it JHEP\/} {\bf 04}, 172 (2014).

\bibitem{Aad:2014kba}
G.~Aad, {\it et~al.\/}, {\it Eur. Phys. J.\/} {\bf C75}, 229 (2015).

\bibitem{Abazov:2014qba}
V.~M. Abazov, {\it et~al.\/}, {\it Phys. Rev.\/} {\bf D90}, 111101 (2014).

\bibitem{Khachatryan:2014iia}
V.~Khachatryan, {\it et~al.\/}, {\it JHEP\/} {\bf 09}, 094 (2014).

\bibitem{Aaboud:2016fzt}
M.~Aaboud, {\it et~al.\/}, {\it Eur. Phys. J.\/} {\bf C77}, 76 (2017).

\bibitem{Aaij:2011yc}
R.~Aaij, {\it et~al.\/}, {\it Phys. Lett.\/} {\bf B707}, 52 (2012).

\bibitem{Aaij:2016bqq}
R.~Aaij, {\it et~al.\/}, {\it JHEP\/} {\bf 06}, 047 (2017). [Erratum:
  JHEP10,068(2017)].

\bibitem{Li:2010hc}
G.~Li, M.~Song, R.-Y. Zhang, W.-G. Ma, {\it Phys. Rev.\/} {\bf D83}, 014001
  (2011).

\bibitem{Mao:2011kf}
M.~Song, W.-G. Ma, G.~Li, R.-Y. Zhang, L.~Guo, {\it JHEP\/} {\bf 02}, 071
  (2011). [Erratum: JHEP12,010(2012)].

\bibitem{Gong:2012ah}
B.~Gong, J.-P. Lansberg, C.~Lorce, J.~Wang, {\it JHEP\/} {\bf 03}, 115 (2013).

\bibitem{Lansberg:2013wva}
J.~P. Lansberg, C.~Lorce, {\it Phys. Lett.\/} {\bf B726}, 218 (2013). [Erratum:
  Phys. Lett.B738,529(2014)].

\bibitem{Lansberg:2013qka}
J.-P. Lansberg, H.-S. Shao, {\it Phys. Rev. Lett.\/} {\bf 111}, 122001 (2013).

\bibitem{He:2015qya}
Z.-G. He, B.~A. Kniehl, {\it Phys. Rev. Lett.\/} {\bf 115}, 022002 (2015).

\bibitem{Sun:2014gca}
L.-P. Sun, H.~Han, K.-T. Chao, {\it Phys. Rev.\/} {\bf D94}, 074033 (2016).

\bibitem{Lansberg:2017chq}
J.-P. Lansberg, H.-S. Shao, N.~Yamanaka, {\it Phys. Lett.\/} {\bf B781}, 485
  (2018).

\bibitem{Lansberg:2016rcx}
J.-P. Lansberg, H.-S. Shao, {\it JHEP\/} {\bf 10}, 153 (2016).

\bibitem{Lansberg:2006dh}
J.~P. Lansberg, {\it Int. J. Mod. Phys.\/} {\bf A21}, 3857 (2006).

\bibitem{Andronic:2015wma}
A.~Andronic, {\it et~al.\/}, {\it Eur. Phys. J.\/} {\bf C76}, 107 (2016).

\bibitem{Alwall:2014hca}
J.~Alwall, {\it et~al.\/}, {\it JHEP\/} {\bf 07}, 079 (2014).

\bibitem{Shao:2012iz}
H.-S. Shao, {\it Comput. Phys. Commun.\/} {\bf 184}, 2562 (2013).

\bibitem{Shao:2015vga}
H.-S. Shao, {\it Comput. Phys. Commun.\/} {\bf 198}, 238 (2016).

\end{thebibliography}

\end{document}